\RequirePackage{etoolbox}
\csdef{input@path}{%
 {sty/}
 {img/}
}%
\csgdef{bibdir}{bib/}

\documentclass[generic]{imsart}

\usepackage{amsthm}
\usepackage{mathrsfs}
\usepackage{mathtools}
\usepackage{amsmath,amssymb}
\usepackage{natbib}
\usepackage[colorlinks,citecolor=blue,urlcolor=blue,filecolor=blue,backref=page]{hyperref}
\usepackage{graphicx}
\usepackage{todonotes}
\usepackage{graphicx}
\usepackage{stix}
\usepackage{algorithm,algorithmic}
\usepackage{booktabs}
\usepackage{rotating} 
\usepackage{microtype}
\usepackage{array, threeparttable, booktabs,  caption}
\usepackage{xcolor}

\startlocaldefs
\numberwithin{equation}{section}
\theoremstyle{plain}

\endlocaldefs

\newcommand{\given} { \,|\, }
\newcommand{\pr}[1]{ \Pr {\left[#1\right]} }
\newcommand{\KL}[2] { \mathrm{KL} {\left(#1 \, \| \, #2\right)} }
\newcommand{\SN}{\mathrm{SN}}
\newcommand{\SP}{\mathrm{SP}}
\newcommand{\TP}{\mathrm{TP}}
\newcommand{\FN}{\mathrm{FN}}
\newcommand{\FP}{\mathrm{FP}}
\newcommand{\TN}{\mathrm{TN}}
\newcommand{\MCC}{\mathrm{MCC}}

\begin{document}

\begin{frontmatter}
\title{Bayesian Estimation of Gaussian Graphical Models with Predictive Covariance Selection}
\begin{aug}
\author{\fnms{Donald R.}\snm{Williams}\thanksref{addr1}},
\author{\fnms{Juho} \snm{Piironen}\thanksref{addr2}},
\author{\fnms{Aki} \snm{Vehtari}\thanksref{addr2}},
\and
\author{\fnms{Philippe} \snm{Rast}\thanksref{addr1}}
\runauthor{D. Williams et al.}
\runtitle{Bayesian Predictive Covariance Selection}
\address[addr1]{Department of Psychology, University of California, Davis, USA}
\address[addr2]{Department of Computer Science, Aalto University, Espoo, Finland}
\end{aug}
%
\begin{abstract}
Gaussian graphical models are used for determining conditional relationships between variables. This is accomplished by identifying off-diagonal elements in the inverse-covariance matrix that are non-zero. When the ratio of variables ($p$) to observations ($n$) approaches one, the maximum likelihood estimator of the covariance matrix becomes unstable and requires shrinkage estimation. Whereas several classical (frequentist) methods have been introduced to address this issue, fully Bayesian methods remain relatively uncommon in practice and methodological literatures. Here we introduce a Bayesian method for estimating sparse matrices, in which conditional relationships are determined with projection predictive selection. With this method, that uses Kullback-Leibler divergence and cross-validation for neighborhood selection, we reconstruct the inverse-covariance matrix in both low and high-dimensional settings. Through simulation and applied examples, we characterized performance compared to several Bayesian methods and the graphical lasso, in addition to TIGER that similarly estimates the inverse-covariance matrix with regression. Our results demonstrate that projection predictive selection not only has superior performance compared to selecting the most probable model and Bayesian model averaging, particularly for high-dimensional data, but also compared to the 
the Bayesian and classical glasso methods. Further, we show that estimating the inverse-covariance matrix with multiple regression is often more accurate, with respect to various loss functions, and efficient than direct estimation. In low-dimensional settings, we demonstrate that projection predictive selection also provides competitive performance. We have implemented the projection predictive method for covariance selection in the R package \href{https://github.com/donaldRwilliams/GGMprojpred/blob/master/README.md}{\textbf{GGMprojpred}} .



\end{abstract}
\begin{keyword}
\kwd{Bayesian}
\kwd{Gaussian graphical models}
\kwd{Neighborhood selection}
\kwd{Projection predictive selection}
\kwd{KL-divergence}
\kwd{Cross-validation}
\kwd{Sparsity}
\end{keyword}

\end{frontmatter}
\section{Introduction}
\label{sec:Intro}
Gaussian graphical models (GGMs) are used for \textit{covariance selection}, in which  conditional dependencies between random variables are characterized \citep{ Dempster1972, Peng2008}. This is accomplished by identifying off-diagonal elements in the inverse-covariance matrix (i.e., precision matrix) that are non-zero. When these covariances are standardized and the sign reversed, they correspond to partial correlations. Assuming multivariate normality, partial correlations imply pairwise conditional dependence (i.e., direct effects) while controlling for all other variables included in the model \citep{Baba2004, Baba2005}. This stands in contrast to marginal correlations, where the associations include indirect and overlapping effects, thereby limiting their use in assessing unique relationships among a set of variables \citep{Schafer2005,Roverato2017}. That GGMs characterize conditional relationships among random variables has led to extensive use across the sciences, in both methodological and applied contexts. 
For example, GGMs are commonly used to characterize genetic co-expression networks \citep{Wang2016}, functional connectivity between brain regions \citep{Das2017}, and more recently to understand chronic mental illnesses \citep{McNally2015MentalSystems}.

The partial correlation matrix $\tilde{\textbf{\textit{P}}} = (\tilde{\rho}_{ij})$ can be obtained by 
inverting the sample covariance matrix $\Sigma$, where $\Sigma^{-1}$ is the corresponding precision matrix $\Omega$. Here, with the covariances $\hat{\omega}_{ij}$ and precisions $\omega_{ii,jj}$ $\in$ $\Omega$,
the partial correlations $\rho_{ij}$ are computed as
\begin{equation} \label{eq:1.1}
\tilde{\rho}_{ij} =\frac{ - \omega_{ij}}{\sqrt{\omega_{ii}\omega_{jj}}},\hspace{.15cm}\text{where}\hspace{.15cm}\Omega = 
\begin{bmatrix} 
\omega_{ii} &                   \\
  \vdots          &  \ddots            \\
\omega_{ij} &   \cdots   & \omega_{jj} \\
\end{bmatrix}
= \Sigma^{-1}.
\end{equation}
In ideal situations, for example when the ratio of variables $p$ to observations $n$ is sufficiently small ($p/n \ll 1$), the customary maximum likelihood (ML) $\frac{1}{N} \sum_{i = 1}^{n}  (X_i - \bar{X}) (X_i - \bar{X})^\top$ can provide an accurate estimate for $\Sigma$. Hence, in this case, inversion to obtain $\Omega$ is straightforward. However, in many practical applications, there are often many variables relative to the number of observations. In these situations, ML becomes unstable and less accurate as $p \rightarrow n$ which is further magnified when inverting $\Sigma$ \citep{Won2009,Ledoit2004}. This does not only influence accuracy of estimation, but also the ability to correctly identify non-zero elements. For example, while confidence intervals for the precision matrix have nominal coverage when $p/n \ll 1$ \citep{Drton2004,Williams2018b}, they were shown to have less than nominal probabilities even when the $p/n$ ratio was below 0.50 \citep{Jankova2015}. These errors are a result of increased variability in the eigenvalues $\lambda$  \citep{Kuismin2016}, from which the determinant of the matrix is computed: 
\begin{equation}
\label{eq:det}
\text{det}(\Sigma) = \prod_{i = 1}^p \lambda_{i},\hspace{0.05 cm} i \hspace{0.05 cm} \in    \{1,...,p\}.
\end{equation}
Further, since the number of non-zero eigenvalues is min($n$, $p$) \citep{Kuismin2017}, it is clear from Equation \ref{eq:det} that $\Sigma$ is singular det($\Sigma$) = 0 when $n < p$. As a result, in this situation, it is not possible to compute the ML estimate for $\Omega$ \citep{Hartlap2007}.


To address these issues, in the classical literature (i.e., frequentist), two primary lines of research have emerged to estimate $\Omega$. The first surrounds accurate estimation with convex combination estimators \citep{Kuismin2017}, and not necessarily determining non-zero relationships \citep{Ledoit2004,Ledoit2004a}. Of course, once obtained, a decision rule can be adopted for identifying non-zero relationships \citep{Huang2013a,Kubokawa2008,Schafer2005}. Although this typically requires some form of null hypothesis testing for the partial correlations, for example using the approach introduced in \citet{Schafer2005a}, recently \citet{Kuismin2016} applied a stepwise strategy directly to $\omega_{ij}$ $\in$ $\Omega$. The most popular approaches \citep{Wang2012, Yuan2013AEstimation}, however, optimize the penalized maximum likelihood with respect to a $\ell$-1 norm constraint. This approach not only shrinks $\omega_{ij}$ towards zero, but can also set the elements to exactly zero ($\omega_{ij} = 0$). 
There are two strategies for utilizing $\ell$-1 regularization \citep{Kramer2009,Friedman2008}, the first of which estimates $\Omega$ directly such as the popular graphical lasso (glasso) method \citep{Friedman2008}, whereas the second avoids the issues of directly estimating $\Omega$ altogether. Instead, the latter uses a column-wise multiple regression strategy. Here, assuming  the input $n \times p$ matrix  is standardized, each variable $p$ is regressed against the remaining $p - 1$ variables. The partial correlation then follows
\begin{equation}  \label{eq:1.2}
\rho_{ij} = \text{sign}(\beta_{i}^{(j)})
\sqrt{\beta_{i}^{(j)}\beta_{j}^{(i)}},
\end{equation}
where $\beta_{i}^{(j)}$ is the penalized coefficient for predictor $X_{j}$ and response $X_{i}$. Several regression approaches, including ridge \citep{Kramer2009}, partial least squares \citep{Tenenhaus2008}, and lasso  have been used in this context \citep{Meinshausen2006}. In this work, we introduce a Bayesian method that also utilizes a regression based approach for both accurate estimation and determining conditional relationships ($\omega_{ij} \neq 0$) in high dimensional settings $n < p$.

Our primary focus is on fully Bayesian estimation of GGMs, which remain relatively uncommon in practice compared to classical methods. While several classical regression strategies have been employed, only recently was a Bayesian lasso regression characterized \citep{Li2017}. Perhaps the most noteworthy Bayesian approaches makes use of the $G$-Wishart distribution, which is conjugate for $\Omega$ whose covariances are constrained to be zero \citep{Atay-Kayis2005,Dobra2010BayesianData}. The earliest approaches utilized reversible jump MCMC \citep{Green1995}, for example in \citet{Wang2012} and  \citet{Lenkoski2013}, whereas the most recent work use a birth and death MCMC algorithm \citep{Mohammadi2015BDgraph:Models}. Both provide a familiar framework for Bayesian model selection, in that covariance selection uses an indicator function analogous to customary stochastic search algorithms for general linear models \citep{George1993,OHara2009}. That is, the underling conditional independence structure can be obtained with either Bayesian model averaging or maximum a posterior probability. Alternative Bayesian approaches, on the other hand, determine non-zero relationships with credible intervals. Recently, block-wise sampling algorithms have been developed to estimate Bayesian versions of the graphical lasso \citep{Wang2012, Khondker2013}, in addition to an adaptive version described in \citet{Wang2012}. 

Although these Bayesian methods have compared favorably to classical methods, there are several important limitations that we address in this work. First, as noted above, the ML estimate and corresponding intervals are adequate when $p/n \ll 1$. However, most of the simulation studies have almost exclusively characterized performance in low-dimensional settings \citep{Mohammadi2015a}. To our knowledge, performance of $G$-Wishart distributions have not been characterized in high-dimensional settings. Second, since Bayesian approaches provide a measure of parameter uncertainty via the posterior distribution, parameters are never actually zero and covariance selection must be achieved with an explicit decision rule. In \citet{Khondker2013}, for example, they used used credible intervals that ranged from 10-50 \%, none of which had a clear justification. Further, in high dimensional settings \citep{VanderPas2016}, credible intervals are unlikely to have nominal frequentist properties that are known to translate into performance measures for estimating GGMs \citep[i.e., specificity = 1 - $\alpha$]{Williams2018b}.

In the present paper, we introduce a novel Bayesian method for estimating Gaussian graphical models. Our method does not rely on posterior model probabilities or credible interval width, but we adopt a decision theoretic perspective. To determine the conditional independence structure, we use projection predictive selection, described in more detail in section~\ref{sec:PPM}, that allows for variable exclusion ($\omega_{ij} = 0$) based on predictive utility \citep{Goutis1998,Dupuis2003}. In the context of the general linear model, this approach was shown to outperform information criteria, cross-validation scores, or selecting the most probable model \citep{Piironen2017}. We also focus on high dimensional settings ($n < p$), where the ML estimate does not exist, but also examine consistency in typical asymptotic settings (fixed $p$; $n \rightarrow \infty$). This not only shows the generality of the proposed method, but also provides a contrast to the known inconsistencies of glasso \citep{Leppa-aho2017a,Williams2018a,Kuismin2016}. 


The remainder of the manuscript is organized as follows. Thus far we have been exclusively discussing covariance selection in terms of $\Omega$, so we first introduce the customary notation and terminology for GGMs (section \ref{sec:GGM}), and then describe the relationship between column-wise multiple regression and the elements of $\Omega$. Second, we then outline the projection predictive method in the context of both low-dimensional and high-dimensional GGMs. For the former, posterior samples are generated with the Bayesian bootstrap, whereas for the latter we use the horseshoe prior distribution (section \ref{sec:PPM}). In section~\ref{sec:Simulation}, we use numerical experiments to evaluate our methods performance compared to $G$-Wishart BMA and MAP estimation, in addition to the Bayesian graphical lasso. As a point of reference, we also include the classical glasso and an alternative $\ell_1$-regularization method that also constructs $\Omega$ with multiple regression. In the  following section, we use each method to estimate a known genetic regulatory network. We end by discussing limitations and future directions.


\section{Gaussian Graphical Models}
\label{sec:GGM}

We now introduce notation and terminology specific to graphical models. Depending on the field, undirected graphical models refer to covariance selection models and random Markov fields. Here we adopted the term Gaussian graphical model, because it is common in the Bayesian literature and provides an informative description of the method. For example, consider a \textit{p}-dimensional random variable $X$ that follows a multivariate normal distribution 
\begin{equation}  \label{eq:2.1}
X = \{X_1, ... , X_p\} \sim \mathcal{N}(\mu, \Sigma), 
\end{equation}
where $\Sigma$ is an unknown, but positive definite covariance matrix and $\mu$ is a vector of means. The inverse of the covariance matrix is $\Omega = \Sigma^{-1}$, where the $i$th row and $j$th column is denoted by $\Omega_{ij}$. Without loss of generality, we assume that all variables have been standardized to have mean zero and variance one: 
\begin{equation}
0 = \{\mu_1, ..., \mu_p\}^\mathsf{T}, \hspace{.15 cm} \text{and} \hspace{.15 cm} (\Sigma_{ii}) = 1. 
\end{equation} 
Expanded to multiple observations, with $n$ independent samples, let \textbf{\textit{X}} be the $n \times p$ data matrix. The graph is then denoted by $\mathcal{G} = (V, E)$. Here $V$ = $\{1,..., p\}$ is the node set and $E \subset V \times V$ is the edge set. Thus, while $V$ represents the total number of columns in \textbf{\textit{X}}, each dimension of Equation \ref{eq:2.1} is denoted by $\textit{X}_{i}$. The edge set for $\mathcal{G}$ contains nodes $(X_{i}, X_{j})$ that share a conditional relationship $X_i \not\Vbar X_j | \boldsymbol{X}_{V- i,-j}$. In contrast, conditionally independent nodes $X_i \Vbar X_j | \boldsymbol{X}_{V-i,-j}$  are not included in $E$ and correspond to the zero elements within $\Omega$ ($\omega_{ij} = 0$). The maximum number of edges possible is $\frac{V(V-1)}{2}$, which corresponds to the number of non-zero off-diagonal elements $\omega_{ij} \in$ $\Omega$.

\subsection{Regression-Based Interpretation}
We estimate $\mathcal{G}$ by fitting node-wise regression models, a procedure known as neighborhood selection. The random variables can be partitioned into two groups
$X = (X_i, X_{-i})$, where $X_{-i} = (X_1,\ldots, X_{i - 1}, X_{i + 1},\ldots, X_p)$. Here, assuming $X$ follows a multivariate normal distribution (Equation \ref{eq:1.2}), the conditional distribution of $X_i$ given $X_{-i}$ is also normally distributed \cite{Anderson2003}, hence,
 
\begin{equation}
X_i | X_{-i} \sim \mathcal{N}(\mu_i + \Sigma_{i,-i} \Sigma_{-i,-i}^{-1} (X_{-i} - \mu_{-i}),\Sigma_{ii} - \Sigma_{i, -i} \Sigma_{-i, - i}^{-1}\Sigma_{-i, i}).
\end{equation}
The equivalent regression equation is denoted as
\begin{equation}
\label{eq:reg_inv}
X_i = \alpha_i + X_{-i}^{\top}\beta_{(i)} + \varepsilon_i,
\end{equation}
where $\alpha = \mu_i - \Sigma_{i,-i} \Sigma_{-i,-i}^{-1} \mu_{-i}$, $\beta_{(i)} = \Sigma_{-i, -i}^{-1} \Sigma_{-i,i}$ is a $p - 1$ dimensional vector, and $\varepsilon_{i} \sim \mathcal{N}(0, \Sigma_{ii} - \Sigma_{i,-i} \Sigma_{-i, -i}^{-1} \Sigma{-i,i})$ is independent of $X_i$ \citep{Yuan2010HighProgramming, Li2017a}. The inverse, resulting in the $i$th column of $\Omega$, follows

\begin{align}
\omega_{ii} &= (\Sigma_{ii} - \Sigma_{i,-i} \Sigma_{-i,-i}^{-1}\Sigma_{-i,i})^{-1}, \\   \nonumber
&= \text{Var}(\varepsilon_k)^{-1}, \\ 
\Omega_{-i, i} &= -(\Sigma_{ii} - \Sigma_{i,-i} \Sigma_{-i,-i}^{-1}\Sigma_{-i,i})^{-1} \Sigma_{-i, -i}^{-1} \Sigma_{-i, i}, \\ \nonumber
&= - \text{Var}(\varepsilon_k)^{-1} \beta_{(i)},
\end{align}
where the resulting precision matrix is

\begin{equation}
\label{eq:inv_mat}
\Omega = 
\begin{bmatrix}
\frac{1}{\text{Var}(\epsilon_k)}_{11} & \frac{- \beta_{12}}{\text{Var}(\epsilon_k)}_{11} &  \cdots & \frac{- \beta_{1p}}{\text{Var}(\epsilon_k)}_{11}   \\
\frac{- \beta_{21}}{\text{Var}(\epsilon_k)}_{22} & \frac{1}{\text{Var}(\epsilon_k)}_{22} & 
\cdots &  \frac{- \beta_{2p}}{\text{Var}(\epsilon_k)}_{22} \\


 \vdots &  \vdots & \ddots  & \vdots \\


\frac{- \beta_{p1}}{\text{Var}(\epsilon_k)}_{pp} & \frac{- \beta_{p2}}{\text{Var}(\epsilon_k)}_{pp} & 
\cdots &  \frac{1}{\text{Var}(\epsilon_k)}_{pp} \\

\end{bmatrix}.
\end{equation}
In the GGM literature, it has been shown that this approach approximates $\Omega$. However, when using least squares, it should be noted that the relationship is exact only when $\Sigma$ can be inverted ($p < n$). Detailed analytic proofs, that we use to derive $\Omega$ in high-dimensional settings (section \ref{sec:symmetry}), are provided in \citet{Stephens2018}.
This is a fundamental problem and in the following section we present a solution for cases where $p > n$.

\section{Projection Predictive Selection}
\label{sec:PPM}
Projective model selection is a model simplification technique and fundamentally a two-step procedure that provides a decision theoretically correct inference after selection. That is, the full posterior is projected to the restricted subspace instead of forming the usual posterior given constraints. Importantly, the projection method is general in that selection only requires samples from the posterior, thus allowing for flexible model specification. 

\subsection{Model Fitting}
The first step requires the construction of the encompassing reference model $M_*$, defined as
\begin{equation}
\label{eq:regression}
\mathbf{y} = \boldsymbol{X} \beta \hspace{0.05 cm} + \hspace{0.05 cm} \varepsilon, 
\hspace{0.05 cm} \varepsilon \sim \mathcal{N}(0, \sigma^2),
\end{equation}
where $\textbf{\textit{X}}$ is a $n \times p - 1$ matrix that includes a subset of nodes $X_{- i}$ from $V$ $\{1,...,p\}$, $\beta$ is a $p-1$ dimensional vector including $\beta_j^{(i)} = \{j,...,p - 1\}$, and \textbf{y} is the $i$th node to be predicted. This model best captures our assumptions and uncertainties related to the problem, including the choice of prior distributions for the task at hand (e.g., the dimensions of the data). 

Building upon Equation \ref{eq:regression}, for the purpose of regularization, we use the horseshoe prior distribution
\begin{align}
\label{eq:horseshoe}
\beta_j |\lambda_{j}, \tau &\sim \mathcal{N}(0, \lambda_{j}^2\tau^2),\hspace{.15 cm} j \in \{1, ..., p - 1 \},   \\ 
\lambda_{j} &\sim  \text{C}^+(0, 1), \nonumber \\
\tau &\sim \text{C}^+(0, \tau_0),   \nonumber 
\end{align}
where $\text{C}^+$ is a half-Cauchy distribution for the local and global hyperparameters denoted with $\lambda$ and $\tau$, respectively. The latter provides shrinkage to all estimates, whereas the local  hyperparameter allows stronger signals to avoid shrinkage, thus making this choice of prior ideal for identifying non-zero relationships. It is possible to incorporate prior expectations in regards to sparsity, denoted $\tau_0$, that serves as the scale for the prior on $\tau$. Described in \citet{Piironen2017}, the value for $\tau_0$ can be computed as
\begin{equation}
\label{eq:node_wise_edges}
\tau_0 = \frac{p_0}{D - p_0} \frac{\sigma}{\sqrt[]{N}},
\end{equation}
where $p_0$ denotes the number of expected edges and $D$ is the number of variables included in the model ($p - 1$). Since priors can have diminishing returns as $n \rightarrow \infty$, the  term $\frac{\sigma}{\sqrt[]{N}}$ allows for scaling with the data. In practice, while it is recommended to include relevant information when available, \citet{Piironen2017a} demonstrate robustness of the results with respect to the specific value. Here we assume a generic value ($\tau_0$ = 1)

\subsection{Neighborhood Selection}
In the second stage, for each of the $p$ regression models, we then determine which nodes can be removed, or set to zero ($\hat{\beta}_{j}^{(i)} = 0$), without introducing considerable predictive loss compared to the original reference model $M_*$ (Equation \ref{eq:regression}). In the predictive framework, the model parameters of the reduced models, herein referred to as submodels including nodes $x$, are determined so that the resulting predictive distribution remains close to that of $M_*$ \citep{Vehtari2012}. A theoretically justified and computationally convenient way to do this is by minimizing the discrepancy between the predictive distributions conditional on the parameter values, where the discrepancy $\delta$ is defined in terms of Kullback-Leibler (KL) divergence averaged over the empirical distribution for $X$ \citep{Goutis1998,Dupuis2003}
\begin{align}
	 \theta_\perp 
	 &= \arg \min_{\theta' \in \Theta_\perp} \delta(\theta,\theta') \nonumber \\
	 &\triangleq  \arg \min_{\theta' \in \Theta_\perp}
	 \frac{1}{n} \sum_{i=1}^n \KL{p(\tilde y \given \theta, X_i, M_*)}{p(\tilde y \given \theta', x_i, M_\perp)}.
\label{eq:proj}
\end{align}

This defines a parameter mapping $\Theta \mapsto \Theta_\perp$ where $\theta \in \Theta$ denotes the parameters of the reference model and $\theta_\perp \in \Theta_\perp$ the parameters of the projected submodel $M_\perp$.
Given a set of posterior draws $\{\theta_s\}_{s=1}^S$ for the reference model, we can then project these individually according to Equation~\eqref{eq:proj} to obtain the projected posterior draws $\{\theta_\perp^s\}_{s=1}^S$ for any submodel $M_\perp$.
It can be shown that the projected regression coefficients are obtained analytically by the least squares solution where the target values 
 are replaced by the fit of the reference model for a particular posterior draw \citep[see the appendix of][]{Piironen2017}. 
The projection loss is then defined as the mean discrepancy over the posterior of the reference model
\begin{align}
 \label{eq:loss}
	L = \frac{1}{S}\sum_{s=1}^S \delta(\theta_s,\theta_\perp^s).
\end{align}
We then seek a reduced model with minimal loss by using a forward search strategy, in which nodes are sequentially added that minimize $\delta$. A direct consequence of defining the model parameters of the submodels as projections of the reference model is that the prior is only specified for the reference model and this information is also then transmitted to the submodels in the projection. 

To assess the accuracy of the reduced models in the second step, we use approximate leave-one-out (LOO) cross-validation with Pareto smoothed importance sampling (PSIS), which avoids the repeated fitting of the reference model, but requires the selection to be performed $n$ times. In (PS)IS-LOO, the posterior draws can be treated as draws from the LOO posteriors after weighting them by the importance weights
\begin{equation}
\label{eq:weights}
 w_{is} \propto \frac{1}{p(y_i \given \theta_s)}
\end{equation}
which are then smoothed to stabilize the LOO estimates \citep{Vehtari2016}. Thus the projection is carried out with the data point $i$ left out, where the submodel draws $\{\theta_s\}_{s=1}^S$ are weighted with $w_{is}$ when predicting the left-out point. This gives an estimate of the predictive accuracy $\hat u_k$ with some standard error $s_k$ for a given model complexity~$k$ (number of nodes, for instance).
As an accuracy measure, we use log predictive density (LPD).

Our decision rule for achieving sparsity ($\hat{\beta}_{i}^{(j)} = 0$) is based on $\hat u_k$. We select the simplest model that has an acceptable difference $\Delta u> 0$ relative to the reference model $\hat u_*$ with some level of confidence $\alpha$. That is, we choose the simplest model satisfying 
\begin{equation}
\label{eq:decision}
\pr{u_*-u_k \le \Delta u} \ge 1-\alpha,
\end{equation}
which is estimated using the Bayesian bootstrap. The choices for $\Delta u$ and $\alpha$ are made 
on subjective grounds depending on how much one is willing to sacrifice predictive accuracy for making the model simpler. In other words, sparsifying $\mathcal{G}$ can be viewed as a trade-off between node-wise predictive loss and parsimony. A reasonable but generic choice for $\Delta u$ would be 5 or 10 percent of the difference $\hat u_* - \hat u_0$, where $\hat u_0$ denotes the accuracy estimate for the simplest possible (null) model \citep{Piironen2017}.

\subsection{Symmetrization}
\label{sec:symmetry}
For each node-wise regression, after the selection is completed, the projected estimates (including zero and non-zero elements) are placed into a $p$ $\times$ $p$ matrix $\Lambda_{i, j \neq i}$. In low-dimensional settings this matrix would be symmetric, with all estimates having the same sign, but in high-dimensional settings this is not guaranteed. We use the ``or-rule'', $\left(\hat{\beta}_{j}^{(i)} \vee \hat{\beta}_{i}^{(j)} \right) \ne 0$, to achieve symmetric non-zero entries in $\Lambda$, in which the neighborhoods are re-projected to include the inconsistent estimate(s). 

\subsubsection{Partial Correlation Matrix}
We then use the approach, described in \citet{Kramer2009}, to compute the partial correlation matrix

\begin{align}
\label{eq:pcor}
\tilde{\textbf{\textit{P}}}_{i, -i} &= [\hat{\rho}_{ij}], \hspace{0.10 cm} j \in \{1,..., p -1\}, \\ \nonumber
 \hat{\rho}_{ij} &= \text{sign}\big(\Lambda\big) \text{min} 
\Big\{1, \sqrt[]{\Lambda_{ij} \hspace{0.05 cm} \circ  \hspace{0.05 cm} \Lambda_{ij}^{\top}     }  \Big\}, \\ \nonumber
&if \hspace{0.10 cm} \text{sign}(\Lambda_{ij}) = \text{sign}(\Lambda_{ji}), \\ \nonumber
&\text{and 0 otherwise}, \\ \nonumber
\end{align}
where $\circ$ denotes the Hadamard product between the corresponding matrix elements. This approach ensures that the partial correlations are bounded [-1, 1], and are well defined in high-dimensional settings \citep{Kramer2009}.

\subsubsection{Precision Matrix}
\label{sec:prec_construct}
Although the proposed method, with Equation \ref{eq:pcor}, provides an estimate for the partial correlations, it is also important to consider an estimate for $\Omega$. In low-dimensional settings, following Equation \ref{eq:inv_mat}, this is straight forward.  We thus follow \citet{Yuan2010HighProgramming} and \citet{Liu2017}, each of which used node-wise regressions to construct $\Omega$ in high-dimensional settings.

The diagonal and off-diagonal elements, from Equation \ref{eq:inv_mat}, are defined as 

\begin{equation}
\omega_{ii} = \frac{1}{\text{Var}(e_k)_{ii}} \hspace{0.10 cm} \text{and} \hspace{0.10 cm} \omega_{ij} = \frac{- \beta_{ij}}{\text{Var}(e_k)_{ii}}
\end{equation} 
where $\text{Var}(e_i)$ is the residual variance. With many variables, due to over-fitting, $\text{Var}(e_i)$ can become very small, thereby produce non-optimal estimates for $\omega_{ii}$ and $\omega_{ij}$. We thus computed the projected residual variance, that included only the selected nodes for each neighborhood, as

\begin{align}
\label{eq:prec_symm}
\hat{\Omega}_{ii}   &=  \frac{1}{\text{Var}(e_k)_{ii}}, \\ \nonumber
\text{Var}(e_k)_{ii} &= y - \Lambda_{i, -i} X_{-i}, \\ \nonumber
\end{align}
where  $y = X_i$ (Equation \ref{eq:reg_inv}), $X_{-i}$ is the model matrix excluding the $i$th variable, and $\Lambda_{i, -i}$ is the matrix containing the projected estimates. We then use Equation \ref{eq:prec_symm}, for all $i \neq j$, to compute the off-diagonal elements 

\begin{equation}
\hat{\Omega}_{ij} = -\frac{1}{2}\Bigg(\frac{\Lambda_{i, i \neq j}}{\text{Var}(e_k)_{ii}} + \frac{\Lambda_{j, j \neq i}}{\text{Var}(e_k)_{jj}} \Bigg),
\end{equation}
where symmetry is obtained by averaging the values. While there are alternative strategies, such as $\tilde{\Omega}_{ij} \leftarrow \text{min} \{\hat{\Omega}_{ij}, \hat{\Omega}_{ji}\} $, both were shown to provide an adequate estimate for $\Omega$ \citep{Cai2011AEstimation}. Whereas this procedure ensures $\Omega$ is asymptotically positive definite \citep{Fan2016}, a correction for finite sample positive definiteness may be needed. We iteratively find the nearest positive definite matrix, as in \citet{Higham2002ComputingFinance} and \citet{Houduo2006AMATRIX}, but with respect to the estimated graphical structure, thereby ensuring the edge set remains intact.

\begin{figure}[!t]
\includegraphics[width = 3.95 in, height = 1.5 in] {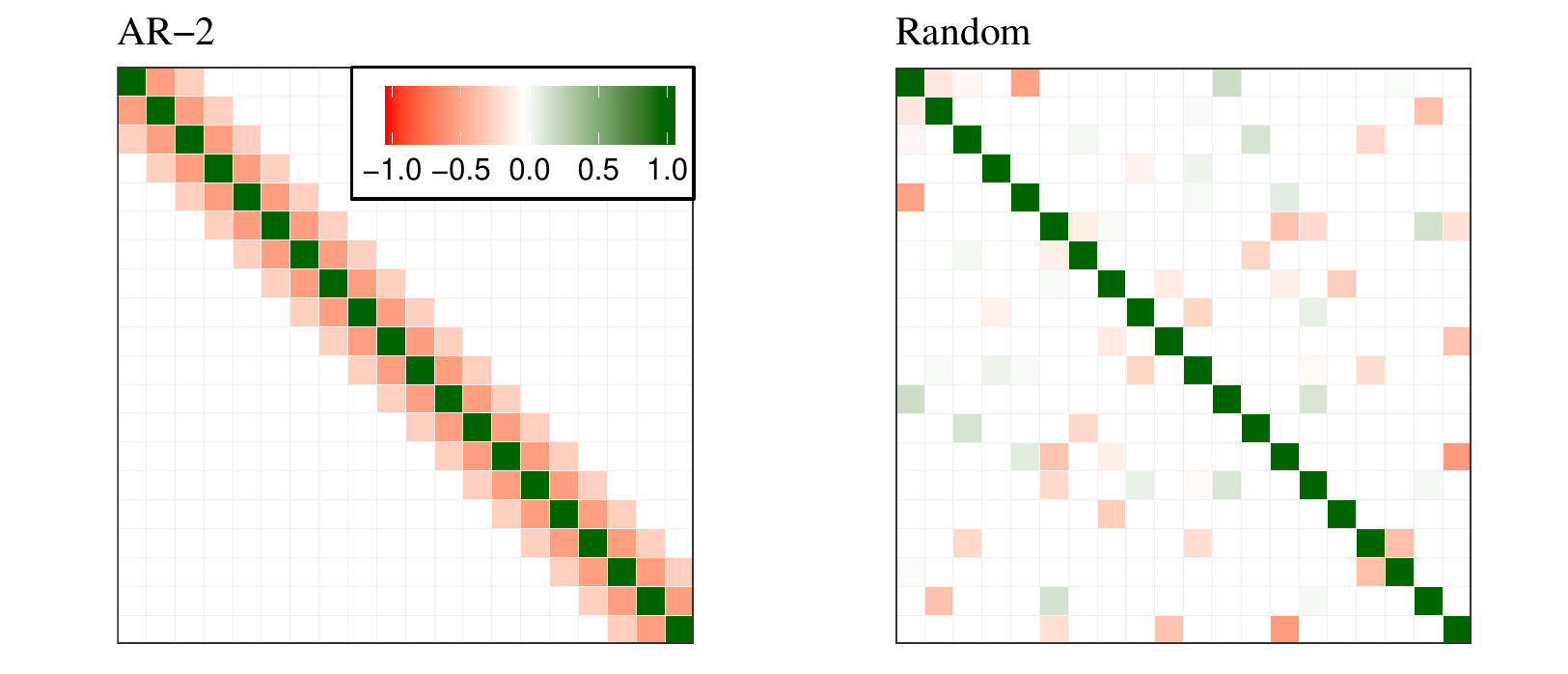}
\caption[size = 6]{Graphical structures used in the low-dimensional simulations (section \ref{sec:low_d}). The precision matrices were standardized, and sign reversed, resulting in the partial correlation matrix. In the random graph, there was 30 \% probability of sharing a connection. }
\label{fig:low_d_graph}
\end{figure}

\section{Numerical Experiments}
\label{sec:Simulation}
In the following, we present a series of experiments in which our aim is to demonstrate the generality of projection predictive \textit{covariance selection}. We first examine model selection consistency in low-dimensional settings, and for high-dimensional settings we also examine various loss functions for $\Omega$. For both dimensions of data, we include the same estimation methods that were specifically chosen to provide an informative contrast to the projection method. The first Bayesian approach, that uses a birth-death MCMC sampler, allowed for obtaining the BMA and MAP graphical structures \citep[R package: \textbf{BDgraph}]{Mohammadi2015BDgraph:Models}. The second Bayesian approach, the graphical lasso (BGL), allowed for comparing the projection method to using credible intervals as the decision rule \citep[low-dimensional: 95 \%; high-dimensional: 50 \%]{Wang2012, Khondker2013}. For the classical methods, we included the customary graphical lasso with EBIC \citep[R package: \textbf{glasso}]{Friedman2008,Foygel2010}, in addition to the TIGER method that estimates $\Omega$ with $\ell_1$-regularized regression \citep[R package: \textbf{flare}]{Liu2017}. For the former we sequenced through 50 potential values for lambda (0.01 to 0.50) and assumed 0.5 for the gamma parameter of EBIC, whereas for the TIGER method lambda was computed as $\sqrt{\text{log}(p) / n}$  \citep{Liu2017}.

For the proposed method, outlined in section \ref{sec:PPM}, we chose generic values for $\tau_0$ = 1 (Equation \ref{eq:horseshoe}), $\Delta u = 10 \%$ (Equation \ref{eq:decision}), and $\alpha = 90 \%$ (Equation \ref{eq:decision}). This choice of $\tau_0$ follows common practice, whereas the values for $\Delta u$ and $\alpha$ were shown to be reasonable choices for variable selection \citep{Piironen2017}. The models were fitted with the R package \href{https://github.com/donaldRwilliams/GGMprojpred/blob/master/README.md}{\textbf{GGMprojpred}}  that estimates high-dimensional regression models with \textbf{horseshoe} and uses \textbf{projpred} for the projection \citep{Paasiniemi207}. For both the projection method and the Bayesian glasso 1,000 samples were drawn from the posterior (excluding a 1,000 iteration warm-up). We used the default settings, as implemented in each package, to fit the remaining models. All computations were done in R version 3.4.2 \citep{RCoreTeam2016R:Computing}.


\begin{figure}[!t]
\hspace{-.8 cm}\includegraphics[width = 4.75 in, height = 4.45 in] {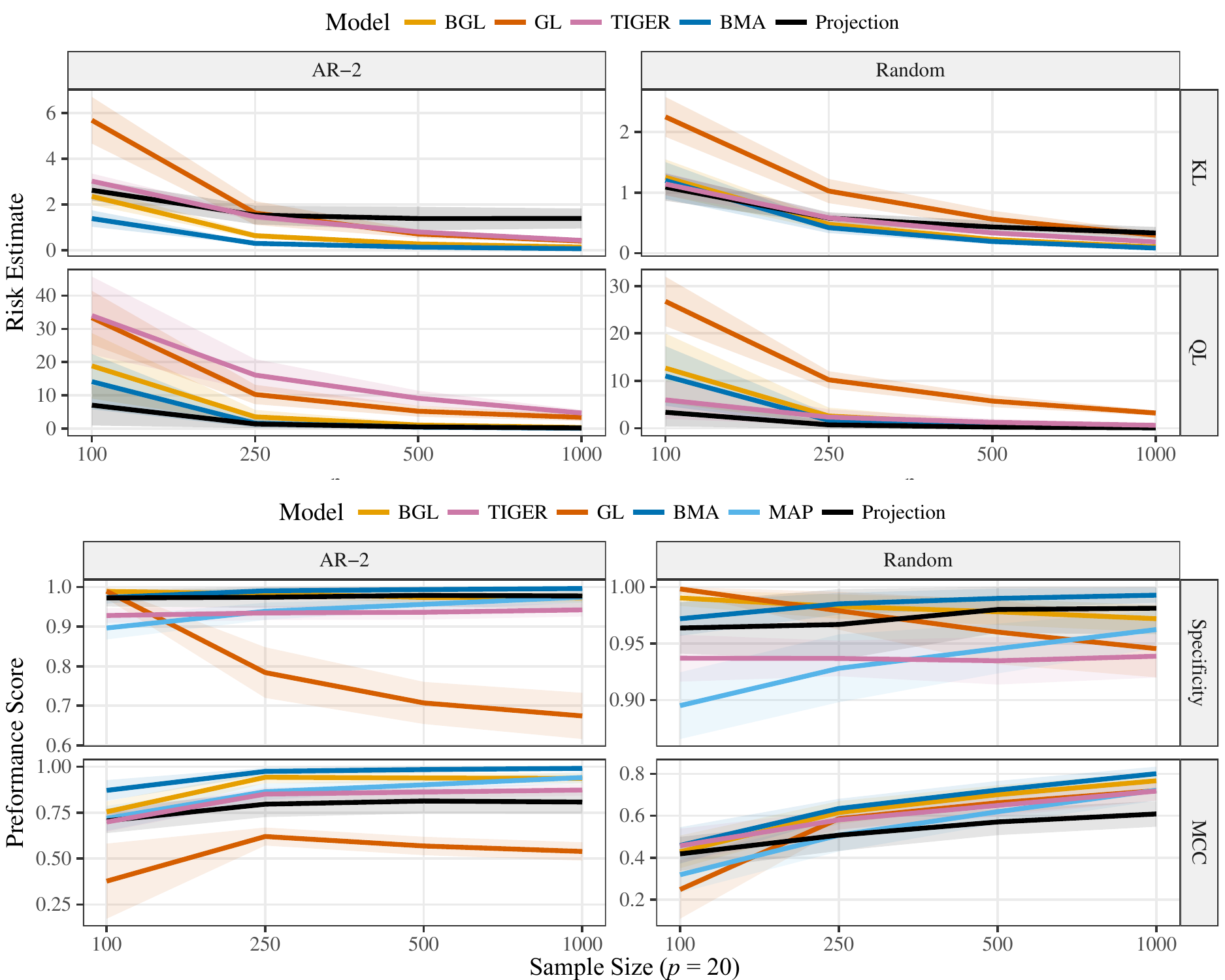}
\caption[size = 6]{Simulation results for the the low-dimensional simulations (section \ref{sec:low_d}).}
\label{fig:results_low_d}
\end{figure}

\subsection{Low-dimensional}
We first examine low-dimensional settings, in the customary asymptotic setting (fixed $p$ = 20), to evaluate consistency as  $n \in \{100, 250, 500, 1000\}$ increased. We considered two graphical structures that are displayed in Figure \ref{fig:low_d_graph}. The first followed an AR-2 process, in which $\omega_{ii} = 1$, $ \omega_{i,i-1} = \omega_{i - 1, i} = 0.5$, and $\omega_{i,i - 2} =  \omega_{i - 2, i} = 0.5$, and $\omega = 0$ otherwise, whereas the second followed a random structure with a 30 \% probability of two nodes sharing a connection. For the latter, the precision matrix was generated from a Wishart distribution $\Omega$ $\sim$ $W_G$($df$ = 20, $I_p$) with 20 degrees of freedom. For each condition, the data were generated from a multivariate normal distribution, with means of zero, and standardized covariance matrices.

Performance was assessed with two loss functions, KL-divergence and quadratic loss, that are defined as
\begin{align}
\text{KL}(\Omega, \hat{\Omega}) &= \text{tr}(\Omega^{-1} \hat{\Omega}) - \text{log}(|\Omega^{-1} \hat{\Omega}|) - p, \\ \nonumber
\text{QL}(\Omega, \hat{\Omega}) &= \text{tr}(\Omega^{-1} \hat{\Omega} - I_p)^2.
\end{align}
We included two measures to assess the accuracy of the graphical structure (edge set identification), including Specificity and the Matthew's correlation coefficient
\begin{equation}
\SP = \frac{\TN}{\TN + \FP} \hspace{0.20 cm} \text{and} \hspace{0.20 cm} \MCC = \frac{\TP \times \TN - \FP \times \FN}{\sqrt[]{(\TP + \FP)(\TP + \FN)(\TN + \FP)(\TN + \FN)}},
\end{equation}
where $\TP$, $\FP$, $\TN$, and $\FN$ are the number of true positives, false positives, true negatives, and false negatives. It should be noted that 1 - SP is the false positive rate, and that the MCC denotes the the correlation between binary variables.

\subsubsection{Results: Risk Estimates}
The results are displayed in Figure \ref{fig:results_low_d}. While all methods were consistent, in that the risk decreased with larger sample sizes, there were some notable differences. The projection method, for both loss functions, provided better performance than BMA. However, for KL-divergence, the Bayesian glasso often had superior performance compared to the projection method. On the other hand, for both graphical structures, the proposed method always had the lowest risk as measured with quadratic loss. For the losses, while performance was competitive with the classical methods, it is important to also consider specificity. For the AR-2 graphical structure, GL performed well with respect to KL-divergence, but specificity consistently declined with larger sample sizes. Importantly, the projection method often had superior performance compared to TIGER, which highlights that our approach for constructing $\Omega$, described in section \ref{sec:prec_construct}, worked well compared to using $\ell_1$-regularized regression to estimate $\Omega$. Further, and interestingly, these results make it clear that directly estimating $\Omega$ does not necessarily provide a more accurate estimate.

\subsubsection{Results: Edge Set Identification}
For edge set identification, it should be noted that the projection method generally had the lowest MCC, except when compared to glasso. However, these results also show that specificity did not decline with larger sample sizes, and that the false positive rate (1- SP) was consistently lower than the classical methods. In fact, specificity was mostly uninfluenced by the sample size, whereas performance varied for the other methods. The alternative Bayesian methods performed well in these performance measures, and in particular BMA and the Bayesian glasso. For the latter, specificity was consistently around 95 \% which suggest the intervals were calibrated with respect to covering zero, thereby suggesting minimal shrinkage. This makes sense, given the dimensions of these data, in that 1 - SP denotes the $\alpha$ level.

\label{sec:low_d}

\subsection{High-dimensional}
We now examine performance in high-dimensional settings, where $n$ was fixed to 50 and $p \in \{25, 50, 100, 150\}$ increased. We considered four additional graphical structures (Figure: \ref{sec:high_d}):

\begin{enumerate}
   \item [1)] \textit{AR-1}: $\Omega^{-1}_{ij}$ = $0.7^{|i - j|}$, and $\Omega^{-1}_{ii}$ = 1.
	\item [2)] \textit{Cluster}: A graph that contains clusters of connections, each of which are  randomly structured graphs. The number maximum number of clusters is max\{2,[$p$/20]\}, with $\Omega \sim W_g(3, I_p)$.
    \item [3)] \textit{Random}: 90 \% of the off-diagonal elements set to zero, and $\Omega \sim W_g(3, I_p)$
    \item [4)] \textit{Scale-free}: A graph generated with the B-A algorithm \citep{Albert2001}, with $p$ - 1      
    edges, and $\Omega \sim W_g(3, I_p)$.
 \end{enumerate}
For each condition, 50 datasets were generated from a multivariate normal distribution, with means of zero, and standardized covariance matrices. It has been noted that methods may perform well in some respects, but not others. We thus considered two more loss functions, the first of which was the $L_2$-loss defined as $L_2(\Omega, \hat{\Omega}) = || \Omega - \hat{\Omega}||_F$. The second loss was mean squared error in which the risk was assessed for the partial correlation matrix. Additionally, we computed sensitivity and the F1-score
\begin{equation}
\SN = \frac{\TP}{\TP + \FN} \hspace{.15 cm} \text{and} \hspace{.15 cm} F1-\text{score} = \frac{\text{2}\TP}{\text{2}\TP + \FP + \FN}.
\end{equation}
The \textit{F1}-score can range between 0 and 1, with 1 denoting perfect identification.

\begin{table}[t]
\centering
\begin{threeparttable}
\caption[size = 5]{Speed of estimation (n = 50) averaged across graphical structures. The parentheses include the standard deviations. Only the results for BMA are displayed (the MAP times are identical). BGL: Bayesian glasso; GL: glasso; TIGER: tuning insensitive graph estimation and regression; BMA: Bayesian model averaging; Projection: projection predictive selection.}
\label{table:1}

\begin{tabular}{cccccccc} \hline
& $p$ & BGL & GL  & TIGER & BMA  & Projection \\ \hline
& 25 & 1.72 (0.10) & 0.04 (0.01)  & 0.19 (0.02) & 3.45 (0.22) & 13.70 (0.24) & \\
& 50 & 7.84 (0.07) & 0.55 (0.28) &  0.29 (0.05) & 14.40 (0.64) &  23.30 (0.55) & \\
& 100 & 52.60 (0.46) & 4.98 (2.81) & 0.72 (0.20) & 104.00 (5.85) & 40.00 (1.22)&\\
& 150 & 129.00 (0.96) & 23.40 (10.90) & 1.50 (0.32) & 345.00 (7.58) & 60.90 (1.13)& \\

\hline
\end{tabular}
\end{threeparttable}
\end{table}

\label{sec:high_d}
\subsubsection{Results: Speed of Estimation}
We first considered the speed of estimation, with the times averaged across graphical structures. These comparisons were non-trivial. To ensure a fair comparison, we implemented each method in the original source code language, but with R serving as a front end. For example, the Bayesian glasso was fitted in Matlab, but was called from the Julia language which was in turn called from R. Not surprisingly the classical methods were much faster, and in particular, the TIGER method that does not perform model selection. On the other hand, glasso had notable variation in the timings which was likely a result of increased estimation time for lower values of the tuning parameter. 

Among the Bayesian methods, the projection method scaled with $p$, which is important for high-dimensional applications. While the projection method was slowest in the lowest dimension ($p = 25$: 13.70 seconds), the time for the highest-dimension setting was the fastest ($p = 150$: 60.90). Indeed, for $p = 150$, the projection method was more than 5 times faster than Bayesian model averaging with a birth-death MCMC algorithm.

\begin{figure}[!t]
\centering
\hspace{- .5 cm}\includegraphics[width = 5 in, height = 1.85 in]{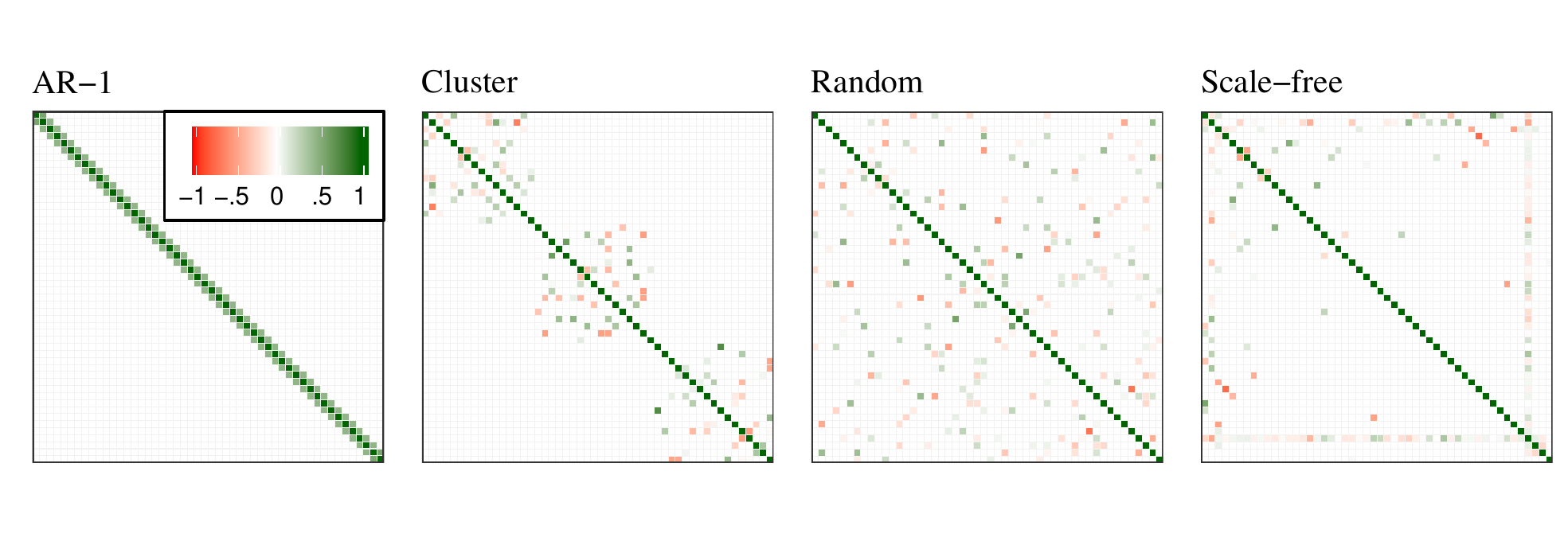}
\caption[size = 4]{\textit{Example graphical structures ($p$ = 50) used in the high-dimensional simulations (section \ref{sec:high_d}). The precision matrices were standardized, and sign reversed, resulting in the partial correlation matrix. In the random graph, for each covariance, there was a 10 \% probability of a conditional relationship.}}
\label{fig:graph}
\end{figure}

\subsubsection{Results: Risk Estimates}
\begin{figure}[!t]
\centering
\hspace{-.80 cm}\includegraphics[width = 5 in, height = 4.45 in] {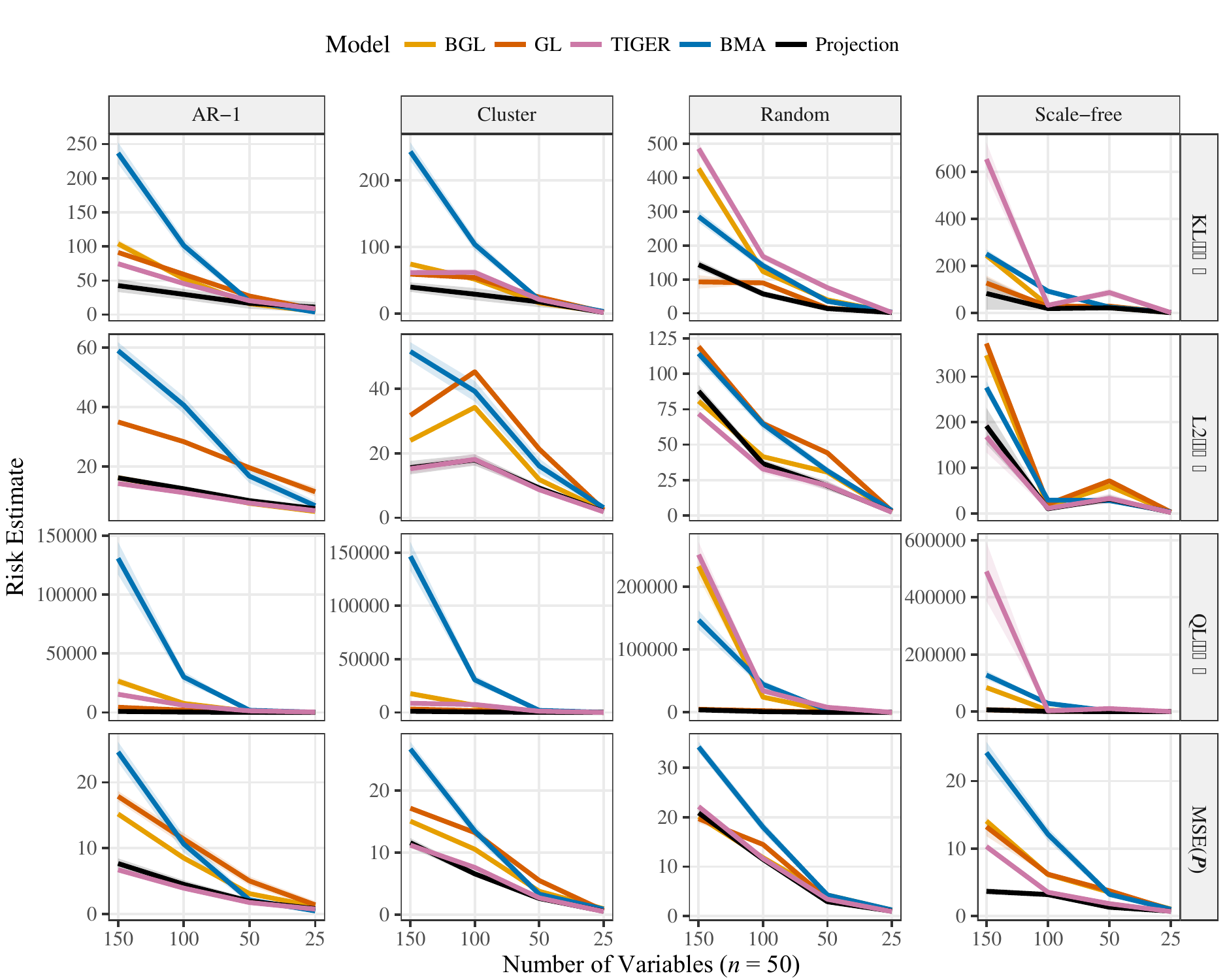}
\caption[size = 4]{\textit{Risk estimates for the high-dimensional simulations (section: \ref{sec:high_d}). GL: glasso; TIGER: tuning insensitive graph estimation and regression; BMA: Bayesian model averaging;  MAP: maximum a posteriori; Projection: projection predictive selection.}}
\label{fig:loss}
\end{figure}

These simulation results are presented in Figure \ref{fig:loss}. We only displayed the BMA results, since they were very similar to the MAP estimates. Most noticeably, the performance of BMA was consistently poor in the high-dimensional setting. In almost all conditions, the risk was highest for the BMA precision matrices ($\Omega$), in addition to the partial correlation matrices (\textit{\textbf{P}}). Specifically, for \textit{\textbf{P}} ($p$ = 150), the MSE was substantially higher than the projection method. For the AR-1 structure, that is, the risk was 4 times greater than for the proposed Bayesian method. Compared to the Bayesian glasso, the projection method also had superior performance in most conditions, particularly in the high-dimensional settings. Of course, it is possible that the decision rule (50 \% credible) could be adjusted to improve performance. We discuss this further in the application section, in addition to the discussion.

The projection method not only provided superior performance compared to the Bayesian methods, but also compared to the classical methods. The TIGER method, which also constructs $\Omega$ from regression, provides an excellent comparison to our method. The risk estimates were strikingly similar for \textit{\textbf{P}}. In fact, these results make it clear that direct estimation of $\Omega$ is not necessarily more accurate. There were some notable difference compared to TIGER, in that the projection method excelled when the measure was either KL-divergence or the quadratic loss. Further, for the scale-free structure, we observed that the TIGER method had increased risk for $\Omega$, in addition to substantial variability. On the other hand, across all conditions, the projection method either had the lowest risk or was competitive, thereby demonstrating the accuracy of our method for constructing $\Omega$ (section: \ref{eq:inv_mat}).

While glasso is one of the most popular methods, it is also the case that more recent methods regularly show superior performance. We thus included glasso as a point of reference, and note the projection method generally had lower risk. In particular, the risk estimates for \textit{\textbf{P}} was often much higher for the glasso method. The most competitive performance was for quadratic loss, although the projection method typically had superior performance.

\subsubsection{Results: Edge Set Identification}
\begin{figure}[!t]
\hspace{-.80 cm}\includegraphics[width = 5 in, height = 4.5 in] {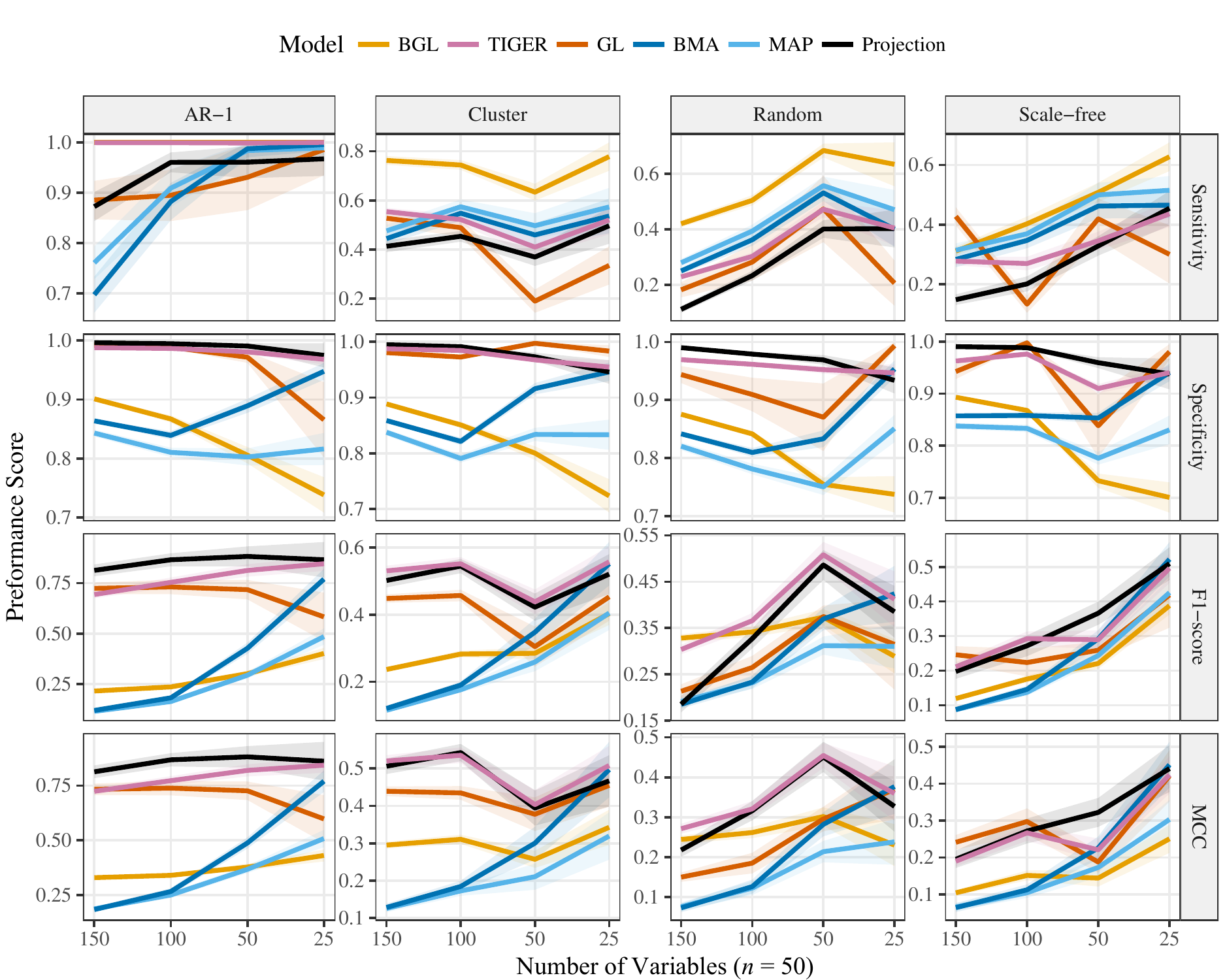}
\caption[size = 4]{\textit{Edge identification for the high-dimensional simulations (section \ref{sec:high_d}). GL: glasso; TIGER: tuning insensitive graph estimation and regression; BMA: Bayesian model averaging;  MAP: maximum a posteriori; Projection: projection predictive selection.}}
\label{fig:sparse}
\end{figure}

 The simulation results are presented in Figure \ref{fig:sparse}. For F1-scores and the MCC, the projection method often had the highest scores. In particular, for the AR-1 and scale-free graphical structures. Interestingly, the performances scores were very similar compared to TIGER which also uses a regression based approach. As such, like the risk estimates, the results indicated that direct estimation of $\Omega$ is not necessarily more accurate.
 
Compared to BMA and MAP estimates, for the high-dimensional settings, the projection method always had superior performance. Notably, the BMA and MAP graphs consistently had F1-scores and MCC's of basically zero ($p$ = 150). This is important, because novel estimation methods are \textit{necessary} for high-dimensional settings, and the BMA and MAP based methods seem most suited for low-dimensional settings. Further, specificity was low for both BMA and MAP (thus the false positive rate was inflated compared to the projection method), which is especially important for high-dimensional graphs in which there are potentially thousands of covariances to consider. While the results for the Bayesian glasso depend on the chosen credible interval (50 \%), the general pattern of results would apply to all decision rules. That is, while specificity approached 90 \% for the high-dimensional settings, it is clear the decision rule would need to change for each dimension of data. That is, specificity consistently diminished with lower dimensions. On the other hand, across most conditions, the projection method had the most consistent performance for specificity, thereby suggesting the generality of our method. However, it should be noted that sensitivity to detect true edges was also lower than the other methods.

These results also make it clear that the projection method compares favorably to glasso. Most notably, the glasso method appears to be erratic (e.g., the scale-free structure) and the most variable. This may be due to performing model selection with EBIC, without ensuring that the assumptions of glasso were satisfied. Within the classical methods, it appears that the TIGER method is a more viable approach for graph estimation.

\begin{figure}[!t]
\hspace{-.50 cm} \includegraphics[width = 5.0 in, height = 3.5 in] {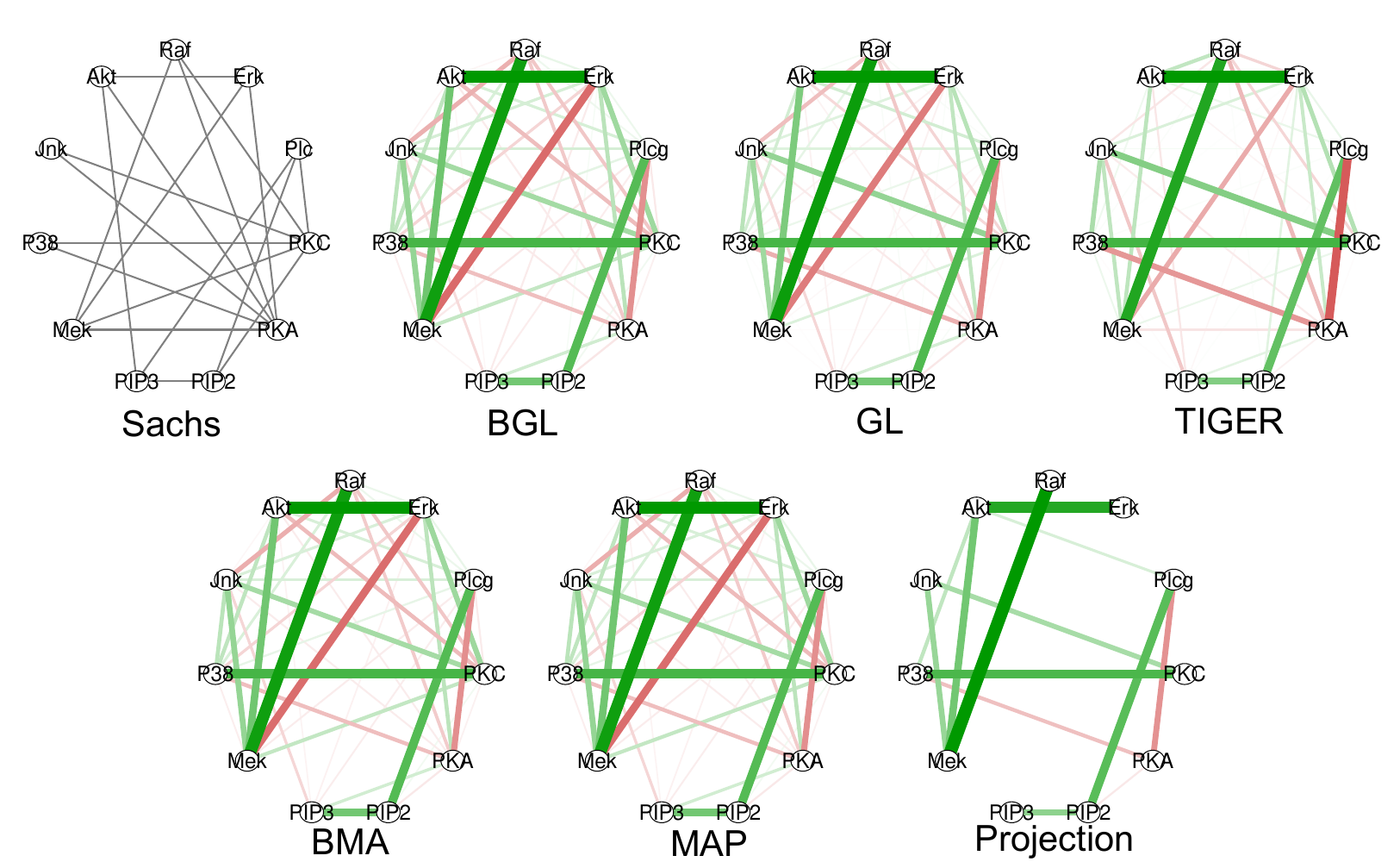}
\caption[size = 4]{\textit{Sachs protein-signaling network. BGL: Bayesian glasso; GL: glasso; TIGER: tuning insensitive graph estimation and regression; BMA: Bayesian model averaging;  MAP: maximum a posteriori; Projection: projection predictive selection.}}
\label{fig:sachs}
\end{figure}

\section{Applications}
\label{sec:sachs}

\subsection{Low-dimensional}
We now compare methods on the \textit{so-called} Sach's network ($p = 11$ and $n = 7,466$). These data were first described in \citet{Sachs2005}, where they used Bayesian methods to suggest causal protein-signaling between phosphorylated proteins and phospholipids. The identified edges were then confirmed with experimental manipulations, thereby providing a unique opportunity to evaluate a known graphical structure (i.e., a ``ground truth").  Because this network is known, it has been used extensively to evaluate the performance of novel GGM methods  \citep{Friedman2008,Khondker2013} or decision rules for existing methods \citep{Kuismin2017}. Based on the present simulations, we used 95 \% credible intervals for the Bayesian glasso.



The estimated graphical structures are displayed in Figure \ref{fig:sachs}, and the performance scores are reported in Table \ref{table:2}. There were clear differences between projection selection and the other methods. Most notably, the alternative methods all estimated very dense graphs in which specificity did not exceed 0.31 (FRP > 0.69). On the other hand, the projection method had the lowest false positive rate but also had the lowest sensitivity. From the estimated graphs, it is clear that the projection method appears to only include edges with strong signals (based on the thickness of the connections). While the F1-scores and the MCC were lowest for the projection method, it was also the case that it was the only method that estimated more true positives than negatives.

Of course, each method has flexibility in model specification, which can possibly improve estimation. We explored this possibility with all three methods. We first examined several $\gamma$ parameters for EBIC model selection with glasso. With a large value ($\gamma = 50$), we were able to decrease the false positive rate from 0.78 to 0.53. These results may reflect that the $\ell_1$-regularized estimates approach maximum likelihood when the sample size becomes large \citep{Kuismin2017}. For the BMA and MAP networks, we made several adjustments to the default settings including (1) more shrinkage, to an identity matrix, by increasing the degrees of freedom parameter $r$ of the Wishart prior distribution (default = 3); and (2) a larger cut-off for the posterior inclusion probability (PIP; default = 0.5). With $r = 100$, the false positive rate decreased by only 0.05 which suggests diminishing returns from the prior distribution in large sample settings. We then set the PIP cut-off to 0.99, and this further reduced the false positive rate from 0.72 to 0.58. We also increased the number of iterations, allowing for more samples from the most probable graphs, but this did not have much of an effect on the false positive rate.  




\begin{table}[t]
\centering
\begin{threeparttable}
\caption{Performance measures for Sach's protein-signaling network. BGL: Bayesian glasso; GL: glasso; TIGER: tuning insensitive graph estimation and regression; BMA: Bayesian model averaging; MAP: maximum a posteriori; Projection: projection predictive selection. FPR: False positive rate.}
\label{table:2}

\begin{tabular}{lcccccc} \hline
            & BGL   & GL   & TIGER& BMA  & MAP   & Projection  \\ \hline
 Sensitivity  & 0.95  & 1.00 & 0.95 & 0.95 & 0.95  & 0.37   \\
 Specificity  & 0.22  & 0.22 & 0.25 & 0.31 & 0.28  & 0.83   \\
 F1-score     & 0.55  & 0.58 & 0.56 & 0.58 & 0.57  & 0.44   \\
 MCC          &  0.22 & 0.30 & 0.24 & 0.29 & 0.27  & 0.23   \\
 FPR    &  0.78 & 0.78 & 0.75 & 0.69 & 0.72  & 0.17  \\

\hline
\end{tabular}
\end{threeparttable}
\end{table}

\begin{figure}[!t]
\hspace{-.50 cm} \includegraphics[width = 5.0 in, height = 3.5 in] {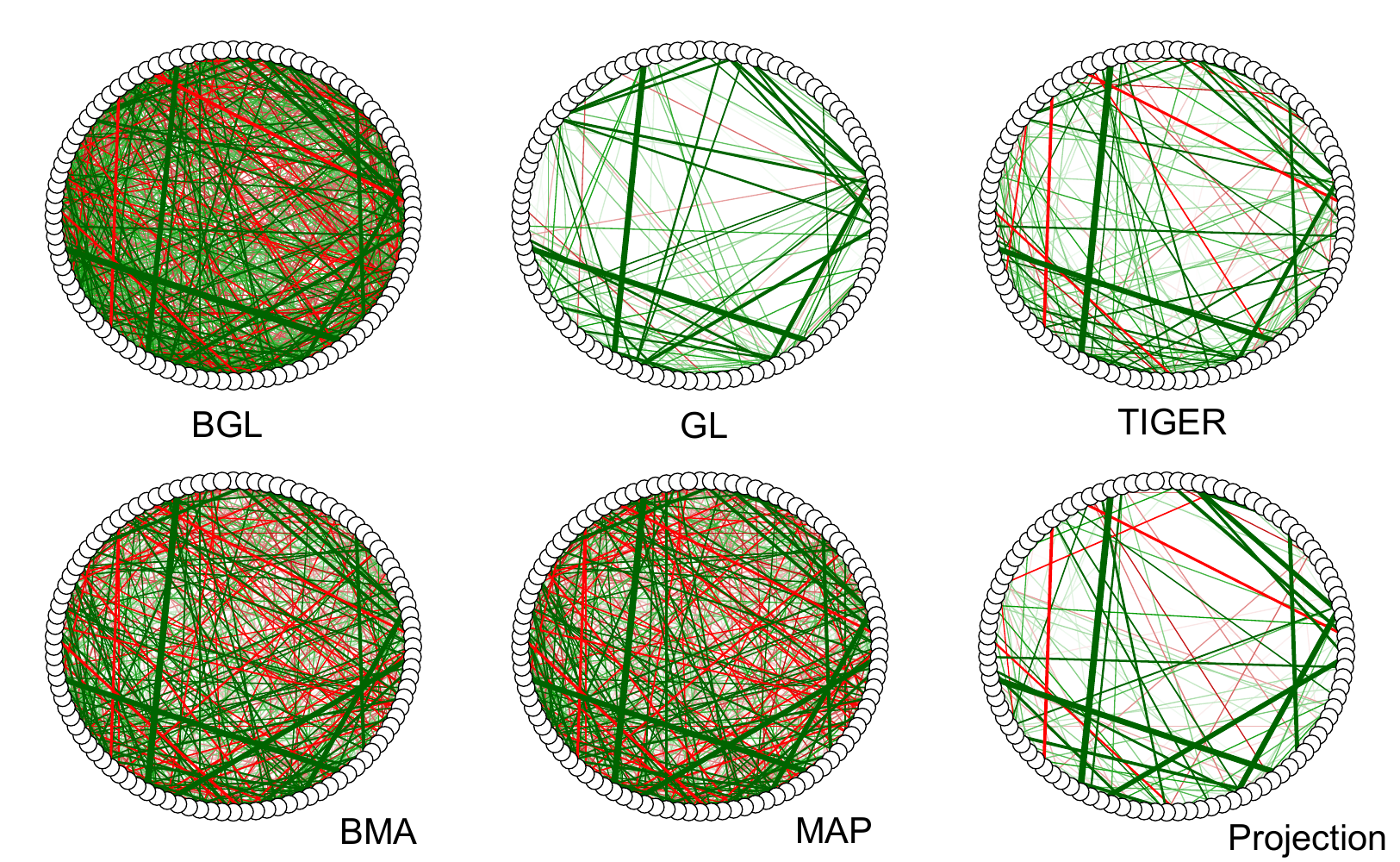}
\caption[size = 4]{\textit{CEU genetic network. BGL: Bayesian glasso; GL: glasso; TIGER: tuning insensitive graph estimation and regression; BMA: Bayesian model averaging; MAP: Maximum a posteriori; Projection: projection predictive selection.}}
\label{fig:ceu}
\end{figure}

\subsection{High-dimensional}
We now analyze a high-dimensional dataset, which includes 100 genes ($p$) from 60 ($n$) unrelated individuals of Northern and Western European ancestry from Utah (CEU). This data is freely available (\href{https://github.com/donaldRwilliams/GGMprojpred/blob/master/README.md}{\textbf{GGMprojpred}}), and thoroughly described in \citep{Bhadra2013JointAnalysis}. Deciding upon the appropriate credible interval, for the Bayesian glasso, was non-trivial. We ultimately chose a 50 \% interval, which was also used in \citet{Khondker2013} and \citet{Li2017a}, but note this can be adjusted to reduce or increase the number of edges.

The estimated graphical structures are displayed in Figure \ref{fig:ceu}. The results correspond closely to the simulation results. That is, BMA (843 edges), MAP (982 edges), and the Bayesian glasso estimated very dense graphs (1,020 edges). We were able to reduce the number of connections, by adjusting the default settings, but the resulting graphs for BMA and MAP remained much denser than the projection method (145 edges). In fact, the proposed Bayesian method provided the sparsest solution, although the glasso (157 edges) and TIGER (242 edges) methods were similar.  In high-dimensional settings, with potentially thousands of partial correlations, this provides a more interpretable graph. 

\section{Conclusion}
We introduced a Bayesian method for estimating Gaussian graphical models with projection predictive selection (section \ref{sec:PPM}). For determining conditional relationships, the proposed method is framed in terms of tolerable predictive loss from a reference model. This decision theoretic framework stands in contrast to posterior model probabilities or credible interval exclusion of zero. In contrast to customary neighborhood selection approaches, that estimate the partial correlation or adjacency matrix, our method provides an estimate of the inverse-covariance matrix. With numerical experiments, we demonstrated that projection predictive covariance selection often outperforms classical and Bayesian methods, both in terms of accurate estimation and edge set identification. For specificity (and thus false positives) in particular, projection selection consistently had the best performance and the scores were largely independent of the simulation condition. Further, in the application section, we showed that our approach similarly produced the sparsest graphical structures, thereby providing a more interpretable network in high-dimensional settings.

In the classical literature, substantial effort has gone into developing regression based approaches for covariance selection \citep{Meinshausen2006,Kramer2009,Peng2009}. These methods often outperform direct estimation, and have less restrictive assumptions \citep{Buhlmann2011StatisticsApplications}. In high-dimensional settings, the projection method had consistently lower risk estimates than BMA and MAP. In particular, the projection method excelled with respect to the log-score (KL-divergence) \citet{Hoeting1999BayesianTutorial}. In other words, while BMA is known to minimize the log-score, these simulations showed that projection selection is more accurate in high-dimensional settings. Direct estimation was also not faster than fitting several regression models, which further highlights an additional advantage of the proposed method. Further, while projective covariance selection finds a trade-off between sparsity and accuracy, it should be noted that sensitivity was consistently lower than the alternative methods. This was not unexpected, for instance, if there are correlating features that carry similar information, then the projection tends to select only one (or a few) of these and ignore the rest because this does not introduce considerable loss in predictive accuracy. For this reason, the projection method might not  be ideal when the goal is to find all the true edges, but rather when we want the simplest possible graph that has almost as low KL/QL compared to a denser graphical structure.

Despite these promising findings, none of the methods provided the best performance for all conditions or performance measures. In section \ref{sec:sachs}, for example, we actively pursued a sparse graph but discovery (minimizing false negatives) could be important to consider. We focused on specificity for two reasons: (1) based on the results, we attempted to outperform projection selection by adjusting the settings of the alternative methods under consideration. This strengthened our conclusion that projection selection estimates the fewest false positives; and (2) the classical methods produced such dense graphs that it was not reasonable to increase the density. Of course, while not explored here, the projection method readily allows for denser graphs by adjusting the decision rule for setting coefficients to zero (section \ref{sec:PPM}).


There are several important limitations for the our method based on projection predictive selection. First, our method is best suited for moderately high-dimensional settings which potentially limits its use in ultra high-dimensions. Second, our method does not initially guarantee a positive definite estimate. However, even with a correction, we showed that the projected precision matrix was often more accurate than direct estimation and a classical method that similarly estimates $\Omega$ with multiple regression. Third, our conclusions are restricted to these methods, conditions, graphical structures, and decision rules. For example, in addition to EBIC based tuning parameter selection, glasso can be used with the rotation information criterion, stability approach for regularization selection, and cross-validation. Finally, we did not compare our method to several approximate Bayesian approaches \citep{Leppa-aho2017a}. Our focus was on fully Bayesian methods, whereas approximate methods often provide only an adjacency matrix, thereby limiting the losses that can be evaluated.

The proposed method can be extended in a number of important ways; for example, to handle mixed graphical models (continuous and binary variables), longitudinal or time series data, or those data in which normative assumptions are not warranted. Additionally, the projection can be carried out on the inverse-covariance matrix directly and alternative samplers can be implemented, which would improve efficiency and allow for improved scalability to higher dimensions. These future directions will be addressed in our future work on this topic.



\section*{Acknowledgements}
Research reported in this publication was supported by three funding sources: (1) The National Academies of Sciences, Engineering, and Medicine FORD foundation pre-doctoral fellowship to DRW; (2) The National Science Foundation Graduate Research Fellowship to DRW; and (3) the National Institute On Aging of the National Institutes of Health under Award Number R01AG050720 to PR. The content is solely the responsibility of the authors and does not necessarily represent the official views of the National Academies of Sciences, Engineering, and Medicine, the National Science Foundation, or the National Institutes of Health. 

\bibliographystyle{ba}
\bibliography{Mendeley.bib}

\end{document}